\documentclass[amsmath,amssymb,superscriptaddress,nobalancelastpage,prl,twocolumn]{revtex4-1}

\usepackage{graphicx}
\usepackage{varioref}
\usepackage{xr-hyper}
\usepackage{xcolor}
\usepackage{hyperref}
\hypersetup{colorlinks,linkcolor=blue,urlcolor=blue,citecolor=blue}
\usepackage{ulem}
\usepackage{braket}

\newcommand{\LSCO}{LSCO}
\newcommand{\znLSCO}{Zn-LSCO}
\newcommand{\LSCOov}{LSCO ($x=0.22$)}
\newcommand{\LSCOovTc}{LSCO ($x=0.22$, $T_c=22$~K)}
\newcommand{\LSCONakamae}{LSCO ($x=0.33$)}
\newcommand{\EuLSCOov}{Eu-LSCO ($x=0.21$)}
\newcommand{\EuLSCOovTc}{Eu-LSCO ($x=0.21$, $T_c=15$~K)}
\newcommand{\EuLSCO}{Eu-LSCO}
\newcommand{\NdLSCO}{Nd-LSCO}
\newcommand{\YBCO}{YBa$_2$Cu$_3$O$_{7- \delta}$}
\newcommand{\Tl}{Tl$_{2}$Ba$_{2}$CuO$_{6+\delta}$}

\newcommand{\dz}{$d_{3z^2-r^2}$}
\newcommand{\dx}{$d_{x^2-y^2}$}

\begin{document}
 \author{M.~Horio}
   \affiliation{Physik-Institut, Universit\"{a}t Z\"{u}rich, Winterthurerstrasse 190, CH-8057 Z\"{u}rich, Switzerland}

 \author{K. Hauser}
   \affiliation{Physik-Institut, Universit\"{a}t Z\"{u}rich, Winterthurerstrasse 190, CH-8057 Z\"{u}rich, Switzerland}
   
    \author{Y. Sassa}
  \affiliation{Department of Physics and Astronomy, Uppsala University, SE-75121 Uppsala, Sweden}

    \author{Z. Mingazheva}
   \affiliation{Physik-Institut, Universit\"{a}t Z\"{u}rich, Winterthurerstrasse 190, CH-8057 Z\"{u}rich, Switzerland}
   
       \author{D.~Sutter}
     \affiliation{Physik-Institut, Universit\"{a}t Z\"{u}rich, Winterthurerstrasse 190, CH-8057 Z\"{u}rich, Switzerland}
     
      \author{K. Kramer}
     \affiliation{Physik-Institut, Universit\"{a}t Z\"{u}rich, Winterthurerstrasse 190, CH-8057 Z\"{u}rich, Switzerland}
     
   \author{A. Cook}
  \affiliation{Physik-Institut, Universit\"{a}t Z\"{u}rich, Winterthurerstrasse 190, CH-8057 Z\"{u}rich, Switzerland}

  \author{E.~Nocerino} 
   \affiliation{Department of Applied Physics, KTH Royal Institute of Technology, Electrum 229, SE-16440 Stockholm Kista, Sweden}
  
  \author{O.~K.~Forslund}
 \affiliation{Department of Applied Physics, KTH Royal Institute of Technology, Electrum 229, SE-16440 Stockholm Kista, Sweden}

   \author{O. Tjernberg}
 \affiliation{Department of Applied Physics, KTH Royal Institute of Technology, Electrum 229, SE-16440 Stockholm Kista, Sweden}
  
 \author{M. Kobayashi}
 \affiliation{Swiss Light Source, Paul Scherrer Institut, CH-5232 Villigen PSI, Switzerland}
 
 \author{A. Chikina}
 \affiliation{Swiss Light Source, Paul Scherrer Institut, CH-5232 Villigen PSI, Switzerland}
 
 \author{N. B. M. Schr{\"o}ter}
 \affiliation{Swiss Light Source, Paul Scherrer Institut, CH-5232 Villigen PSI, Switzerland}
 
 \author{J. A. Krieger}
 \affiliation{Laboratory for Muon Spin Spectroscopy, Paul Scherrer Institute, CH-5232 Villigen PSI, Switzerland} \affiliation{Laboratorium f{\"u}r Festk{\"o}rperphysik,  ETH Z{\"u}rich, CH-8093 Z{\"u}rich, Switzerland}
 
 \author{T.~Schmitt}
 \affiliation{Swiss Light Source, Paul Scherrer Institut, CH-5232 Villigen PSI, Switzerland}
 
  \author{V.~N.~Strocov}
 \affiliation{Swiss Light Source, Paul Scherrer Institut, CH-5232 Villigen PSI, Switzerland}
 
    \author{S.~Pyon}
\affiliation{Department of Advanced Materials, University of Tokyo, Kashiwa 277-8561, Japan}
\author{T.~Takayama}
\affiliation{Department of Advanced Materials, University of Tokyo, Kashiwa 277-8561, Japan}
\author{H.~Takagi}
\affiliation{Department of Advanced Materials, University of Tokyo, Kashiwa 277-8561, Japan}
       
    \author{O. J. Lipscombe}
         \affiliation{H. H. Wills Physics Laboratory, University of Bristol, Bristol BS8 1TL, United Kingdom}
         
     \author{S. M. Hayden}
       \affiliation{H. H. Wills Physics Laboratory, University of Bristol, Bristol BS8 1TL, United Kingdom}
     
     \author{M. Ishikado}
       \affiliation{Comprehensive Research Organization for Science and Society (CROSS), Tokai, Ibaraki 319-1106, Japan}
       
     \author{H. Eisaki}
       \affiliation{Electronics and Photonics Research Institute, National Institute of Advanced Industrial Science and Technology, Ibaraki 305-8568, Japan}

  \author{T. Neupert}
  \affiliation{Physik-Institut, Universit\"{a}t Z\"{u}rich, Winterthurerstrasse 190, CH-8057 Z\"{u}rich, Switzerland}

   \author{M. M\aa nsson}
 \affiliation{Department of Applied Physics, KTH Royal Institute of Technology, Electrum 229, SE-16440 Stockholm Kista, Sweden}
  
   \author{C.~E.~Matt}
   \affiliation{Physik-Institut, Universit\"{a}t Z\"{u}rich, Winterthurerstrasse 190, CH-8057 Z\"{u}rich, Switzerland}
   \affiliation{Swiss Light Source, Paul Scherrer Institut, CH-5232 Villigen PSI, Switzerland}
   \affiliation{Department of Physics, Harvard University, Cambridge, MA 02138, USA.}

  \author{J.~Chang}
    \affiliation{Physik-Institut, Universit\"{a}t Z\"{u}rich, Winterthurerstrasse 190, CH-8057 Z\"{u}rich, Switzerland}

\title{Three-Dimensional Fermi Surface of Overdoped La-Based Cuprates}

\begin{abstract}
We present a soft x-ray angle-resolved photoemission spectroscopy study of overdoped high-temperature
superconductors. 
In-plane and out-of-plane components of the Fermi surface are mapped by varying the photoemission angle and the incident photon energy. No $k_z$ dispersion is observed along the nodal direction, whereas a significant antinodal $k_z$ dispersion is identified for La-based cuprates. 
Based on a tight-binding parametrization, 
we discuss the implications for the density of states near the van-Hove singularity. Our results suggest that the large electronic specific heat found in overdoped La$_{2-x}$Sr$_x$CuO$_4$ can not be assigned to the van-Hove singularity alone. We therefore propose quantum criticality induced by a collapsing pseudogap phase as 
a plausible 
explanation for observed enhancement of electronic specific heat.
\end{abstract}

\maketitle
  
The nature of the pseudogap phase 
in high-temperature cuprate superconductors remains an outstanding problem~\cite{NormanAP2005}. 
It has at the same time been associated with different types of broken symmetries~\cite{ZhaoNatPhys2017,FauquePRL2006,DaouNat2010,HashimotoNatPhys2010,XiaPRL2008} and interpreted as a crossover phenomenon with an ill-defined temperature onset~\cite{WilliamsPRB1998}. 
In recent years, a connection between 
the pseudogap collapse as a function of doping and the Fermi level ($E_\mathrm{F}$) crossing 
of the van-Hove singularity (VHS) has been proposed~\cite{BenhabibPRL2015,Doiron-LeyraudNatCom17,MarkiewiczSCIREP2017,WuPRX18}.
In this scenario, the pseudogap exists only on a hole-like Fermi surface (FS) sheet.
In particular, for the La-based cuprates, it was suggested that the pseudogap phase is truncated 
at the doping where the VHS crosses $E_\mathrm{F}$~\cite{Cyr2017}. 
This coincides approximately with a
maximum in the electronic 
specific heat peaks~\cite{MomonoPHYSICA94,MichonarXiv2018}.
Therefore electronic specific heat  enhancement 
could be a signature of quantum criticality due to the vanishing pseudogap phase at the doping $p=p^*$. Or, it could be  explained simply from density-of-states (DOS) enhancement generated by the VHS. The latter scenario is expected to be significant for quasi-two-dimensional band structures~\cite{MarkiewiczPRB2005}. 
 Experimentally, it has thus become 
important to determine the out-of-plane FS structure of La-based cuprates.



\begin{figure*}
 	\begin{center}
 		\includegraphics[width=1\textwidth]{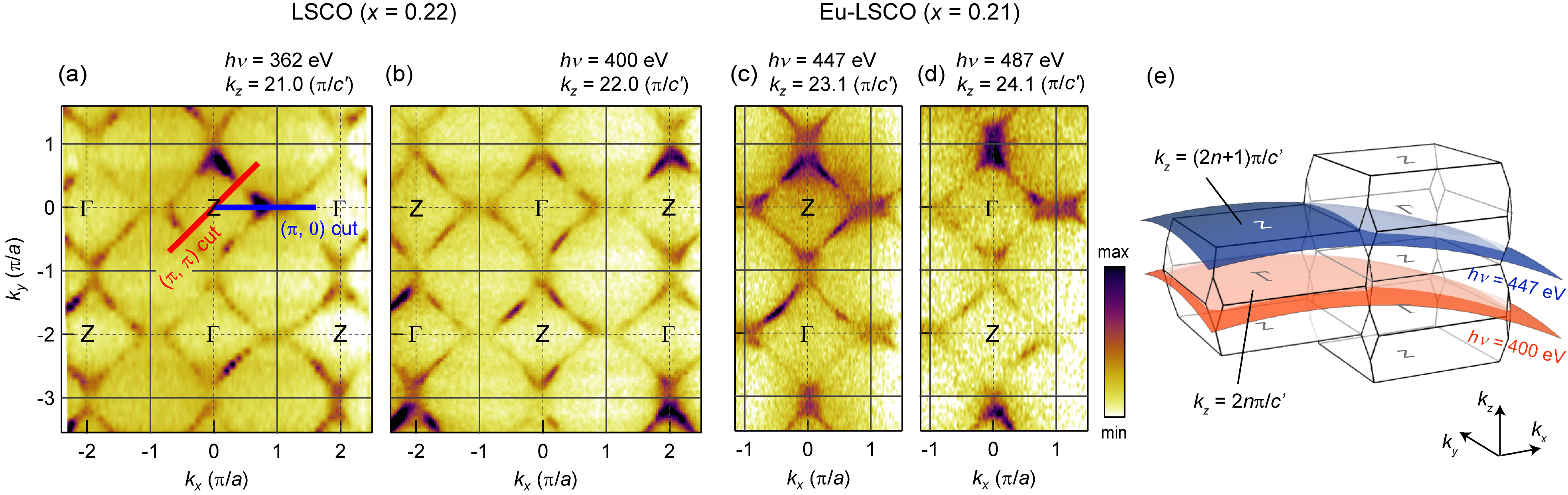}
 	\end{center}
 	\caption{In-plane FSs of  \LSCOov\ (a),(b) and  \EuLSCOov\ (c),(d) measured at $T=12$~K using 
 	SX-ARPES. Photoelectron intensities have been integrated  $\pm$20 meV around $E_\mathrm{F}$. 
 	Corresponding $k_z$ value at $(k_x, k_y)$ = $(0, 0)$ is indicated on the right top of each panel along with the associated incident photon energies. (e) Sketch of the Brillouin zone and location of the in-plane cuts in the 3D momentum space at $h\nu = 400$
 	and $447$ eV. 
 	}	
	 	\label{fig:fig1}
 \end{figure*}

Quantum oscillation (QO) and angle-resolved photoemission spectroscopy (ARPES) experiments are classical 
probes of
the FS structure and quasi-particle renormalization effects~\cite{DamascelliRMP03,SebastianARCMP2015}. Both techniques have been applied 
to overdoped \Tl\ (Tl2201) compounds for which a single large FS sheet 
is observed~\cite{PlatePRL2005,PeetsNJP2006,RourkeNJP10,VignolleNat08}.
The observation of a single QO frequency suggests that if any $k_z$ dependence exists, it is weak.
In contrast, 
angle-dependent magneto-transport experiment has been interpreted as an evidence of a finite FS $k_z$ dispersions~\cite{HusseyNat2003}. To date, ARPES has not provided any
information about three dimensionality of the FS in the cuprates. The vast majority of ARPES experiments have been carried out in the vacuum-ultra-violet 
regime~\cite{DamascelliRMP03}.  For 20--200 eV photons,
the photoelectron mean free path (MFP) is a few \AA~\cite{StrocovJESRP2003}, resulting in considerable $k_z$ broadening~\cite{StrocovJESRP2003}.
Only few ARPES studies of cuprate superconductors exist
in the soft x-ray regime~\cite{ClaessonPRL2004,ClaessonPRB2009,ZabolotnyyPRB2012,MattNatCommun2018}, where much larger MFP and thus better $k_z$ resolution is
reached.
Soft x-ray ARPES (SX-ARPES) has been applied to \YBCO\ \cite{ZabolotnyyPRB2012} to reach bulk sensitivity and overcome the polar catastrophe~\cite{NakagawaNatMater2006}. In La$_{2-x}$Sr$_{x}$CuO$_4$ (\LSCO), the \dz\ band has been probed by use of distinctive photoionization matrix elements in the soft x-ray regime~\cite{MattNatCommun2018}. To date, however, there are no reports on FS $k_z$ dispersion for La-based cuprates. Such information is especially desirable 
since the connection between VHS and pseudogap is most relevant for these compounds~\cite{Doiron-LeyraudNatCom17}.


Here, we apply 
SX-ARPES to reveal the FS $k_z$ dispersion
of three different cuprates, namely, \LSCOov, La$_{1.8-x}$Eu$_{0.2}$Sr$_x$CuO$_4$ (Eu-LSCO, $x=0.21$), and Tl2201.
The first mentioned compound 
does not display any pseudogap physics (i.e. $p>p^*$), and hence the FS is well defined.
Secondly, this specific composition of \LSCO\ has body-centered tetragonal (BCT) crystal structure, and therefore, orthorhombic band folding is avoided~\cite{ChangNJP2008,KingPRL2011}.
No discernible $k_z$ dependence is found along the nodal direction.
By contrast, a clear $k_z$ dependence is found in the antinodal region for the La-based cuprates.
This dispersion is parametrized using 
a single-band tight-binding model. 
Including inter-layer hopping $t_z$ to reproduce the observed band structure,
the corresponding DOS is not large enough to explain the specific heat anomaly.
Our results suggest that the VHS alone cannot account for the specific-heat enhancement, and therefore support the scenario that associates quantum criticality arising from the collapse of the pseudogap phase.

%
Single crystals of \LSCOovTc, \EuLSCOovTc,
 and Tl2201 ($T_c=20$~K)  
were grown by the floating-zone  and self-flux techniques.
Crystal lattice parameters $a$ and $c$ are listed in Table~I.
The sample quality has been demonstrated previously by 
experiments~\cite{LipscombePRL2007,ChangPRB2012,FatuzzoPRB2014,ChangNatComm13,MattPRB2015} 
on the same batch of crystals. Experiments were 
carried out at the 
SX-ARPES endstation~\cite{StrocovJSR2014} of the ADRESS beamline~\cite{StrocovJSR10} at the Swiss Light Source (SLS) of the Paul Scherrer Institute (PSI), 
Switzerland. ARPES spectra were recorded at $T=12$~K with 300--600~eV circularly polarized 
photons 
covering more than three Brillouin zones in both in- and out-of-plane directions. The energy and momentum resolution depend 
on the exact incident photon energy.  
For $500$~eV photons, the 
effective 
resolution is 
$\sim 90$~meV and $\sim 0.02$~$\pi/a$ for energy and momentum, respectively (full width at half maximum).  Measurements were carried out 
with the analyzer slit oriented parallel to the incident x rays as in Ref.~\onlinecite{StrocovJSR2014} .
Pristine surfaces were realized using 
a top post or a cleaving tool~\cite{cleaver}.
To index high-symmetry points in three-dimensional reciprocal 
space $(k_x, k_y, k_z)$, we use the BCT notation with $\Gamma=(0, 0, 0)$, 
Z $=(0, 0, \pi/c')$, $\Sigma = ([1+\eta]\pi/a, 0 , 0)$, and $\Sigma_1 = ([1-\eta]\pi/a, 0 , \pi/c')$,
 where $c'=c/2$ represents CuO$_2$-layer spacing and $\eta = a^2 / 4c'^2$.
The out-of-plane momentum is given by 
\begin{equation}
    \hbar k_z=\sqrt{2m[(h\nu-\phi-E_\mathrm{B}) \cos^2(\theta) +V_0]} + p_c \sin \alpha
\end{equation}
where $m$ is the electron mass, $\phi$ 
the work function, $E_\mathrm{B}$ 
the binding energy, $\theta$ 
the photoemission polar angle, $V_0$ 
the inner potential, $\hbar$ 
the reduced Planck constant, and $p_c$ is the incident photon momentum that is significant 
for SX-ARPES. 
The incident grazing angle $\alpha$ was set to $20^\circ$. For the inner potential, we assumed $V_0=15$~eV consistent with what  has been used for pnictide materials~\cite{XuNatPhys2011,VilmercatiPRB2009}.
For our density-functional-theory (DFT) calculations, 
the WIEN2k package~\cite{Blaha2001} was used.

\begin{figure*}[ht!]
 	\begin{center}
 		\includegraphics[width=0.95\textwidth]{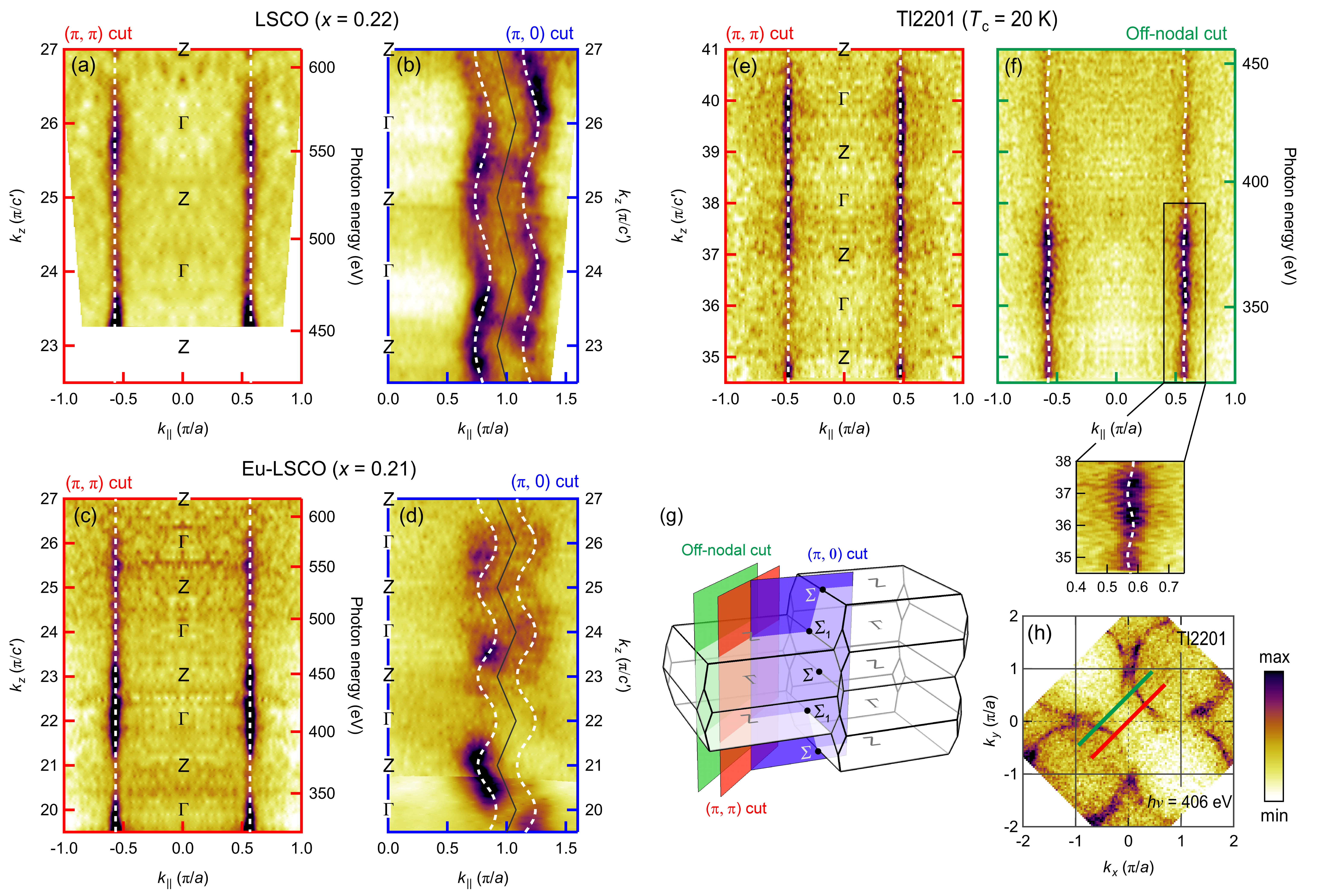}
 	\end{center}
 	\caption{Out-of-plane FS dispersions. (a)--(d) Out-of-plane FS maps  recorded along  nodal ($\pi,\pi$) 
 	and the antinodal ($\pi, 0$) 
 	directions -- 
 	as indicated in (g) -- for \LSCOov\ and \EuLSCOov.
 	(e),(f) $k_z$ dependence of nodal and off-nodal [see (g) and (h)] $k_\mathrm{F}'s$ on the FS of Tl2201. A zoom on the  off-nodal $k_z$ dispersion is provided in (f).  Nodal  out-of-plane maps and data in panels (b) and (f)
 	are symmetrized around $k_{||}=0$. 
 	FSs reproduced by the three-dimensional tight-binding model (see text) are overlaid as white dashed curves. 
 	Photoelectron intensities are integrated  $\pm$20 meV around $E_\mathrm{F}$. Black lines in panels (b) and (d) represent Brillouin zone boundaries. (g) Sketch of the three-dimensional Brillouin zone and location of the cuts. 
 	(h) In-plane FS map of Tl2201 with nodal ($\pi,\pi$) and off-nodal cuts indicated.}	
	 	\label{fig:fig2}
 \end{figure*}

 
Maps of the electron-like~\cite{ChangNatComm13,YoshidaJCMP07,YoshidaPRB2006,RazzoliNJP2010}
FS topology 
of 
\LSCOov\ and \EuLSCOov\ for different values of $k_z$
are shown in Fig.~\ref{fig:fig1}. 
Even though 
$k_z$ varies 
across these maps [Fig.~\ref{fig:fig1}(e)],  strong matrix-element effects complicate
the extraction of any $k_z$ dispersion. 
It is therefore better examined by FS-mapping 
directly along the $k_z$ direction over a wide momentum region. In Figs.~\ref{fig:fig2}(a)--(d), nodal ($\pi$, $\pi$) and  antinodal ($\pi$, 0) cuts as a function of $k_z$ (incident photon energy) 
are shown. In the nodal direction, no discernible dispersion is observed across two different 
Brillouin zones. 
Intensity variations are again assigned to matrix element effects.
Along the antinodal direction, by contrast, a clear dispersion of $k_\mathrm{F}$ is found. The two FS branches separated by the zone boundary disperse $\pi$-shifted along the $k_z$
direction [Figs.~\ref{fig:fig2}(b) and (d)]. This $\pi$-shift is a direct consequence of 
the BCT crystal structure where the $\Gamma$ and the Z points alternate 
in the in-plane direction [Fig.~\ref{fig:fig2}(g)]. As a reference, nodal and off-nodal $k_z$ dispersions of the more two-dimensional Tl2201 system are shown in Figs.~\ref{fig:fig2}(e) and (f).

To parametrize the 
three-dimensional 
FS structure, we use a simple tight-binding model 
decomposed into two terms: $\epsilon_\mathrm{3D}(k_x,k_y,k_z)=\epsilon_\mathrm{2D}(k_x,k_y)+\epsilon_z(k_x,k_y,k_z).$
Although band structure of La-based cuprates involves hybridization of \dx\ and \dz\ orbitals~\cite{MattNatCommun2018}, 
we -- for simplicity --
employ an effective single band  (\dx)
model:
\begin{align}
       &\epsilon_\mathrm{2D}(k_x,k_y) = -\mu + 2t[\cos(k_xa)+\cos(k_ya)]  \\
      &+ 4t'\cos(k_xa)\cos(k_ya) + 2t''[\cos(2k_xa)+\cos(2k_ya)],\nonumber
\end{align}
where $t$, $t'$, and $t''$ represent 
first-, second-, and third- nearest-neighbor
hopping parameters, and $\mu$ is the chemical potential. The out-of-plane dispersion reads:
\begin{align}
    \epsilon_z(k_x,k_y,k_z) =& 2t_{z} \sigma [\cos(k_xa)-\cos(k_ya)]^2 \cos(k_zc') 
\end{align}
where $t_z$ denotes an inter-layer hopping parameter~\cite{MarkiewiczPRB2005}. The term $[\cos(k_xa)-\cos(k_ya)]$ arises from the hybridization between O 2$p$  and Cu 4$s$ or 3\dz\ orbitals 
that promote hopping along the $c$-axis~\cite{AndersenJPCS1995,XiangPRL1996}. A characteristic of the BCT structure is that it has an offset of successive CuO$_2$ planes in the diagonal in-plane direction by ($a$/2, $a$/2), generating an additional factor $\sigma=\cos(k_xa/2) \cos(k_ya/2)$~\cite{MarkiewiczPRB2005}.
The out-of-plane FS of \LSCOov\ 
was fitted to this tight-binding model (Fig.~\ref{fig:fig2}). 
The obtained in-plane hopping parameters (see Table I) are consistent with the previous studies~\cite{YoshidaPRB2006,HashimotoPRB2008}. 
From the $k_z$ antinodal dispersion, we in addition extract the out-of-plane hopping parameter $t_z$. For both \LSCOov\ 
and \EuLSCOov, 
fitting with Eq.~3 provided a good description of the observed dispersion 
with 
$t_z=0.07t$. 
In overdoped La-based cuprates, 
$t_z/t$ thus constitutes a significant fraction. For comparison, overdoped Tl2201, with a hole-like in-plane FS, yields a $k_z$ dispersion [see Fig.~\ref{fig:fig2}(f)] with $t_z/t < 0.015$.

\begin{table}
	\begin{center}
	{\tabcolsep = 3mm
	\begin{tabular}{cccc}
		\hline \hline
			 & LSCO  &  Eu-LSCO & Tl2201 \\  
	    & ($x=0.22$) & ($x=0.21$) & ($T_c = 20$~K) \\ \hline
		$a=b$ & 3.76~\AA  & 3.79~\AA & 3.87~\AA \\
		$c=2c'$ & 13.22~\AA  & 13.14~\AA & 23.20~\AA\\  \hline
		$t'$ & -0.12~$t$  &  -0.14~$t$ & -0.28~$t$ \\
		$t''$ & 0.06~$t$  &  0.07~$t$  &  0.14~$t$ \\
		$\mu$ & 0.93~$t$  &  0.95~$t$  &  1.44~$t$\\
		$t_z$ & 0.07~$t$  &  0.07~$t$  &  ($<$) 0.015~$t$\\ \hline \hline
	\end{tabular}}
	\end{center}
	\caption{Lattice constants and parameters for three-dimensional tight-binding model. $c'=c/2$ represents CuO$_2$-layer spacing. Tight-binding parameters are expressed as a fraction of the nearest-neighbor hopping parameter $t$.
	A fixed ratio $t'' = -0.5t'$ has been assumed~\cite{YoshidaPRB2006}.}
	\label{table:table1}
\end{table}


\begin{figure*}[ht!]
 	\begin{center}
 		\includegraphics[width=1\textwidth]{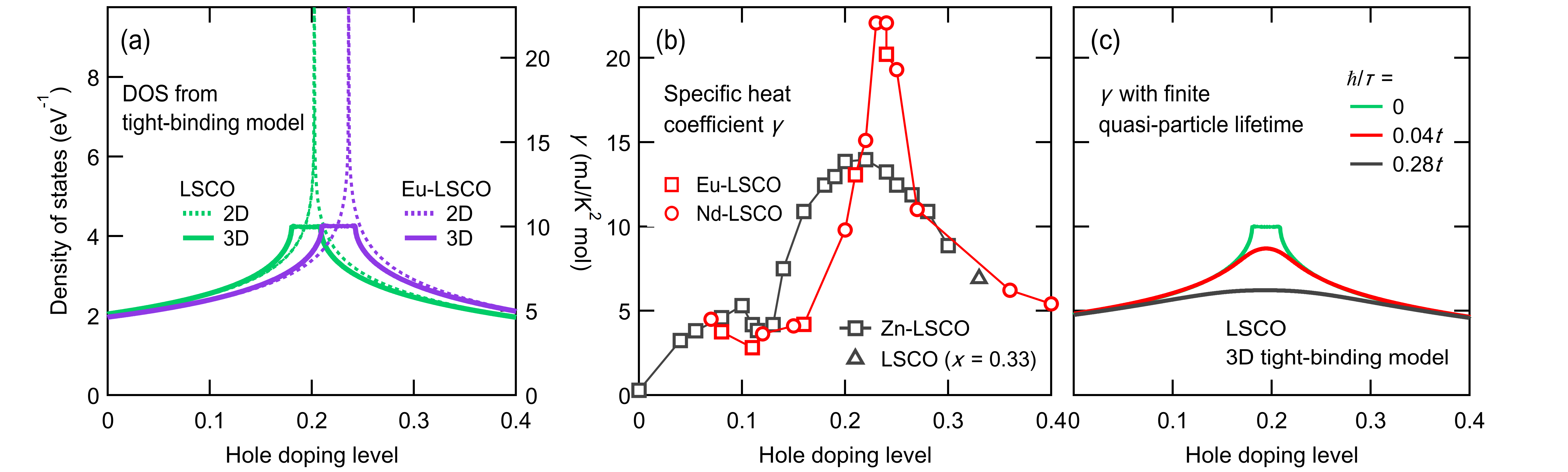}
 	\end{center}
 	\caption{Comparison of calculated DOS with 
 	experimentally extracted Sommerfeld constant $\gamma$.
 	(a) Doping dependence of DOS at $E_\mathrm{F}$ calculated for LSCO and Eu-LSCO from a tight-binding model 
 (see text and Table~I) 
 	with $t_z = 0.07t$ (3D) and $0$ (2D). The diverging peak in two-dimensions is replaced by a plateau in the three-dimensional model. The right axis indicates the electronic specific heat coefficient 
 	$\gamma = \frac{\pi^2}{3} k_\mathrm{B}^2 \times \mathrm{DOS}(E_\mathrm{F})$. (b) Doping dependence of $\gamma$ 
 	reported on \znLSCO~\cite{MomonoPHYSICA94}, \LSCONakamae~\cite{NakamaePRB2003}, \EuLSCO, and \NdLSCO~\cite{MichonarXiv2018} plotted on the same scale as in (a). 
 	(c) $\gamma$ values calculated from the 3D tight-binding model for \LSCO\ with  quasi-particle scattering rates as indicated.}
	 	\label{fig:fig3}
 \end{figure*}

We start our discussion by pointing to a known discrepancy in overdoped \LSCO, 
between bulk hole doping and the FS area~\cite{YoshidaPRB2006,ChangNatComm13}.
The tight-binding extracted 
FS area, equivalent to $p=0.32$ and 0.31 for \LSCOov\ and \EuLSCOov, 
 is significantly larger than the nominal Sr concentrations. 
A stronger $k_z$ dependence in the overdoped region had been put forward as an explanation~\cite{YoshidaPRB2006}. Having measured the three-dimensional FS, 
this scenario is 
eliminated.
It has also been hypothesized that the cleaved surface may have a higher doping than the bulk. Our 
SX-ARPES is more bulk sensitive and should thus alleviate the discrepancy. As this  is not the case,
 this scenario is also not plausible. The filling of overdoped La-based cuprates thus remains puzzling
 but has no qualitative impact on the following discussion.

Having quantified the out-of-plane hopping, it is interesting to discuss transport anisotropy ratios. Sr$_2$RuO$_4$ and overdoped \LSCO\ are isostructural and both display low-temperature Fermi liquid behavior~\cite{NakamaePRB2003,HusseyPRB1998}. The ratio $\rho_c/\rho_{ab}$ between out-of-plane ($\rho_c$) and in-plane ($\rho_{ab}$) resistivity is about 100 for \LSCO~\cite{NakamaePRB2003,NakamuraPRB1993} and 1000 for Sr$_2$RuO$_4$~\cite{HusseyPRB1998} which even has a 
shorter $c$-axis lattice parameter ($c=12.74$~\AA). For overdoped La$_{1.6-x}$Nd$_{0.4}$Sr$_x$CuO$_4$ (\NdLSCO), right at the pseudogap critical doping concentration $p^*=0.24$, an anisotropy factor $\rho_c/\rho_{ab}\sim$ 200
is found~\cite{DaouNatPhys2009,CYRCHOINIERE2010S12}. These values for La-based cuprates are considerably smaller than what has been found in  Tl2201 ($\rho_c/\rho_{ab} \sim$ 1000--2500)~\cite{HusseyPRL1996,KimPRB2004}. 
This is consistent with
first-principle  DFT calculations
that predict $t_z/t=0.12$ for \LSCO~\cite{MarkiewiczPRB2005} and 
0.01 for Tl2201.
For \LSCO, this value of $t_z$ is 
1.7 times larger than what is found by our experiment. 
Assuming for 300--600~eV photons 
a probing depth of 10~\AA, the experimental 
$k_z$ broadening amounts 
to $\sim 0.2 \pi/c'$.
The 
$k_z$ resolution, therefore, does not 
lead to any significant underestimation of $t_z$.
The discrepancy between experiment and DFT calculations is 
thus significant. 
This DFT overestimation of $t_z$ is linked to the \dz\ orbital that influences interlayer hopping. DFT 
places the \dz\ band closer to $E_\mathrm{F}$ than observed 
experimentally in LSCO~\cite{MattNatCommun2018}.
Once the \dz\ band is far from
$E_\mathrm{F}$ as in Tl2201~\cite{HirofumiPRL10}, DFT predicts a $k_z$ dispersion consistent with the experiment.

From the antinodal FS $k_z$ dispersion of LSCO and Eu-LSCO, 
the DOS anomaly generated by the VHS can be estimated. For a given binding energy $\omega$, the two-dimensional DOS$(\omega) = \frac{a^2}{2\pi^2} \frac{dA}{d\omega}$ where $A$ is the constant-energy-surface area. 
The in-plane nearest-neighbor hopping parameter $t=-0.19$ eV is set by the observed nodal Fermi velocity~\cite{ZhouNat2003,YoshidaJCMP07,ChangNatComm13}.  Averaging along the $k_z$ axis yields the DOS($E_\mathrm{F}$) 
versus doping (filling),  Fig.~\ref{fig:fig3}(a), for (i) a two-dimension FS and (ii) the experimentally determined three-dimensional FS.
The divergent peak in the 2D-model is  replaced by a plateau once $k_z$ dispersion is introduced.
The plateau indicates the doping range for which 
the FS character (electron- or hole-like) changes as 
a function of $k_z$. Crystal symmetry imposes two VHS points (separating electron- and hole-like FSs) to exist at $E_\mathrm{F}$ between the $\Sigma$ and the $\Sigma_1$ points.
Irrespective of the splitting along $k_z$ of these VHS points, the DOS remains constant because of the fixed number of singularities. The plateau width and height are primarily set by $\bar{t_z}=t_z/t$
and $1/\bar{t_z}$, respectively. In-plane hopping parameters $t'/t$ and $t''/t$ are less important and experimentally known to vary little with doping~\cite{YoshidaPRB2006}. Due to the weak doping dependence of lattice parameters,
we thus assumed all hopping parameters to be
constants~\cite{MarkiewiczPRB2005}.

The DOS is proportional to the electronic specific heat
(Sommerfeld) coefficient $\gamma = \frac{\pi^2}{3} k_\mathrm{B}^2 \times \mathrm{DOS}(E_\mathrm{F})$ and hence 
directly comparable 
to 
measurements of \znLSCO~\cite{MomonoPHYSICA94}, \LSCONakamae~\cite{NakamaePRB2003}, \EuLSCO, and \NdLSCO~\cite{MichonarXiv2018}
[see Figs.~\ref{fig:fig3}(a) and (b)].
Taking into account the observed $k_z$ dispersion yields a Sommerfeld constant around the VHS that is 0.5--0.7
of the experimental value~\cite{MomonoPHYSICA94,NakamaePRB2003}. 
Including disorder in our evaluation of DOS only enhance the discrepancy [Fig.~\ref{fig:fig3}(c)] 
because finite quasi-particle lifetime $\tau$
suppresses the VHS.
For \znLSCO\ (\EuLSCO\ and \NdLSCO), a 
scattering rate of $\hbar/\tau = 0.28t$ ($0.04t$) is expected~\cite{YoshidaPRB2009,VerretPRB2017,MichonarXiv2018}. In this case
the simulated $\gamma$ peak accounts for less than half of the measured value.
Therefore, the VHS alone cannot 
account for the strong enhancement of the specific heat near $p = p^*$. 
This implies that sources going beyond band structure are required to explain the specific heat of overdoped \LSCO. Quantum criticality originating from the collapse of the pseudogap phase is thus 
a tangible explanation for 
the electronic specific heat enhancement.
In summary, we have revealed the full three dimensional FS structure of overdoped La-based cuprates
using the 
SX-ARPES technique. A significant $k_z$ dispersion was observed on the antinodal FS portion while the nodal part of the FS is non-dispersive. The three-dimensional FS was parametrized using the single-band tight-binding model. In this manner, the out-of-plane hopping term is quantified. Finally, it was shown that the three-dimensional FS structure cannot account for the large electronic specific heat observed in overdoped \LSCO. Quantum criticality emerging from the pseudogap collapse provides 
a plausible explanation for the specific heat anomaly.

%
\begin{acknowledgments}
We acknowledge fruitful discussions with L.~Taillefer, S.~Verret, and A.-M.~S.~Tremblay. 
M.H., D.S., K.K., J.A.K., and J.C. acknowledge support by the Swiss National Science Foundation (grant numbers: CRSII2{\_}160765, PP00P2{\_}150573, BSSGI0{\_}155873, and 200021{\_}165910). Y.S. and E.N. are funded by the Swedish Research Council (VR) with a Starting Grant (Dnr. 2017-05078) and the Swedish Foundation for Strategic Research (SSF) within the Swedish national graduate school in neutron scattering (SwedNess). O.K.F. and M.M. are supported by Marie Sk{\l}odowska-Curie Action, International Career Grant through the European Commission and Swedish Research Council (VR), Grant No. INCA-2014-6426, the Carl Tryggers Foundation for Scientific Research (CTS-16:324), and a VR neutron project grant (BIFROST, Dnr. 2016-06955). T. N. acknowledges support from the Swiss National Science Foundation (grant number: 200021{\_}169061) and from the European Union's Horizon 2020 research and innovation program (ERC-StG-Neupert-757867-PARATOP). The present work was partially supported by JSPS KAKENHI Grant Number JP15K17712. This work was also supported by 
the Knut and Alice Wallenberg foundation. Sample characterizations on Tl2201 were performed by using SQUID magnetometer (MPMS, Quantum Design Inc.) at the CROSS user laboratories. ARPES measurements were performed at the ADRESS beamline of the Swiss Light Source at the Paul Scherrer Institute.
\end{acknowledgments}

%


\end{document}